# Cluster quantum computer on the basis of quasi-particles

V. K. Voronov

Irkutsk State Technical University, 83 Lermontov Str., 664074, Irkutsk, Russia, voronov@istu.edu.ru

The present paper deals with the possibility of creation of the quantum computer in which the role of q-bits is played by quasi-particles. In such a computer, the elementary computation block should represent a cluster created on the basis of the paramagnetic molecules. The latter form heterogeneous spin states in the cluster owing to the presence of interelectron correlations.

Keywords: Quantum computer; strongly correlated systems; entangled states.

The quantum computer is a device which work should be described by parameters, typical for macroobjects. At the same time, the computation procedure in such a device is based on the application of microobjects which are in certain quantum states. According to modern notions, the case in point is quantum particles for which entangled states, detected on macrolevel, are possible. Importantly, such microparticles should be much enough in number that the quantum computer could solve the real tasks /1/. However it is easy to notice that the conditions indicated, in their turn, generate problems in the description of systems containing great number of quantum particles. In this connection, one might specify, at least, two problems.

The first problem is due to the necessity of generation of entangled states. Initially, in the first years of investigations, this problem seemed to be solved by the employment of atom nuclei for this purpose. Some advances have been made in this direction. In particular, the elementary quantum processor on the basis of the phenomenon of nuclear magnetic resonance (NMR) has been created /2,3/. However, a short time later, some basic difficulties have appeared here. They are related to the realization of the entangled states (unambiguously detected on macrolevel) of system nuclear spins containing large number of non-equivalent atoms with respect to chemical environment /4/. It turned out that selective action on a specific nucleus inevitably produces the undesirable effect on states of other resonating nuclei. In this line, it has been proposed /5/ to use the system of q-bits, which would be in a superposition (relative to spin) state by the moment of action of the NMR impulse. It is pertinent to note that this system should attain such a state due to the evolution which is not connected with NMR action planned.

The second problem can be considered as a consequence of the properties caused by collective behavior of quantum particles participating in the computation, in particular, by interelectron correlations if to steal about the system of electrons. The account of effects induced by this behavior is necessary, for example, to describe the properties of strongly correlated systems /6/. In these compounds, several degrees of freedom (spin, charge, spin-orbital, and photon) compete and interact. The investigations conducted over the last twenty years have shown that various physical phenomena are inherent for the compounds containing atoms with unoccupied $3d$-, $4f$- and $5f$-shells. In solid states, such atoms preserve completely or partially localized magnetic moments. A strong interaction of electrons with each other or with collectivized electrons of outer shells is a peculiarity which imparts unique properties to a series of compounds containing the atoms of transitive and rare-earth elements.

If the interelectron correlations are essential elements of behavior of system with many quantum particles, it would be rational to try to use them for the realization of quantum computation procedure. In the work /7/, it has been shown that heterogeneous spin states may exist in heterospin systems, i.e. in coordination compounds of the paramagnetic ions incorporating molecules bearing non-paired electrons as ligands. Since the magnetic and spin moments are interrelated, it is possible to say that in the molecules of the above-mentioned compounds, heterogeneous magnetic states can appear owing to the interelectron correlations.

The appearance of these heterogeneous spin states makes it possible the electrons, localized in these states for some time, to be detected. For a certain time, an electron is placed to one of such states (for example, on $d$-orbital of the central ion). Owing to a specificity of intramolecular spin interactions, the electron is delocalized for some time into other state. Therefrom it is transferred again to the orbitals of metal. Then the cycle repeats. Thus, the state of entanglement, needed for the work of the quantum computer, is realized. Such state can preserve for any time, if the following condition is fulfilled:

$$f(x) = ax - bx^n, \qquad (1)$$

where $a$ and $b$ - constant positive coefficients, $n \geq 2$. If $x \ll 1$, then $bx^n \ll ax$, therefore

$$f(x) \approx ax. \qquad (2)$$

Thus, in the case of (2), $f(x)$ increases linearly with the growth of $x$. If the size $x$ is comparable with 1, then it is impossible to neglect $bx^n$ member, i.e. one should use the initial equation (1) for the description of behavior of the system. Hence, the growth of function deviation at the expense of the $ax$ member will lead to nonlinear restriction owing to subtraction of the $bx^n$ value. Under certain values of $x$, the function $f(x)$ will be close again to zero (or to any specific value) and starts from the beginning. The system will regulate itself as though automatically, because its properties depend on a current state (in this case – from the value of $x$).

As it is known, the description of collective behavior of a system consisting of many quantum particles is often associated with a concept of quasi-particle. For example, the notions on magnetic polaron or ferrone appeared to be fruitful for the description of different heterogeneous charge and spin states in manganites /8/. The theory of interpolar interactions, in particular, the theory of bipolarons of the big radius is often successfully used for the study of interelectron correlations (see /9/ and the references therein). In one of the applications of the theory mentioned, the energy of two polarons is expressed as follows /9/:

$$E = E_1 - J \vec{S}_1 \vec{S}_2, \qquad (3)$$

where $\vec{S}_1$ and $\vec{S}_2$ - are spins of first and second polarons. Important in this expression is the presence of second member, playing the role of the second item in expression (1). Parameter $J$ characterizes the energy of exchange interaction of polarons which, in the long run, depends on an electron and spatial structure, and also on intramolecular dynamics of the corresponding molecular system.

All this allows one to conclude that taking into account interelectron correlations, it is possible to generate the entangled states of q-bits using strongly correlated. Here, the systems of q-bits should be created with a help of the notions on quasi-particles. Therefore, one can speak about the quantum computer, where the quasi-particles play the role of q-bits. The elementary processor of such a quantum computer should have the following structure. This is a cluster composed from paramagnetic molecules. Inside of this cluster there are heterogeneous magnetic states originated from interelectron correlations. Migration of the electron between various states (spin heterogeneities) represents the process of self-organization which will provide the realization (continuous in time) of states entanglement needed for the operation of the quantum computer. The real carriers of entanglement appear to be quasi-particles. By changing the sizes of a cluster as well as its structure (electron and spatial), one can achieve the necessary quantity of quasi-particles in this cluster, and also the degree of their interaction. It is worthwhile to note that the criti-cal radius of a polaron and bipolaron in the ammonia cluster, for example, equals 36 Å and 80 Å, respectively /9/. Current development of physics and technologies of artificial nanostructures enables to create the clusters discussed here of necessary sizes.